# Precision Guided Approach to Mitigate Data Poisoning Attacks in Federated Learning

K Naveen Kumar
Computer Science and Engineering
Indian Institute of Technology
Hyderabad, India
cs19m20p000001@iith.ac.in

C Krishna Mohan
Computer Science and Engineering
Indian Institute of Technology
Hyderabad, India
ckm@cse.iith.ac.in

Aravind Machiry
Electrical and Computer Engineering
Purdue University
USA
amachiry@purdue.edu

## ABSTRACT

Federated Learning (FL) is a collaborative learning paradigm enabling participants to collectively train a shared machine learning model while preserving the privacy of their sensitive data. Nevertheless, the inherent decentralized and data-opaque characteristics of FL render its susceptibility to data poisoning attacks. These attacks introduce malformed or malicious inputs during local model training, subsequently influencing the global model and resulting in erroneous predictions. Current FL defense strategies against data poisoning attacks either involve a trade-off between accuracy and robustness or necessitate the presence of a uniformly distributed root dataset at the server. To overcome these limitations, we present FedZZ, which harnesses a zone-based deviating update (ZBDU) mechanism to effectively counter data poisoning attacks in FL. The ZBDU approach identifies the clusters of benign clients whose collective updates exhibit notable deviations from those of malicious clients engaged in data poisoning attack. Further, we introduce a precision-guided methodology that actively characterizes these client clusters (zones), which in turn aids in recognizing and discarding malicious updates at the server. Our evaluation of FedZZ across two widely recognized datasets: CIFAR10 and EMNIST, demonstrate its efficacy in mitigating data poisoning attacks, surpassing the performance of prevailing state-of-the-art methodologies in both single and multi-client attack scenarios and varying attack volumes. Notably, FedZZ also functions as a robust client selection strategy, even in highly non-IID and attack-free scenarios. Moreover, in the face of escalating poisoning rates, the model accuracy attained by FedZZ displays superior resilience compared to existing techniques. For instance, when confronted with a 50% presence of malicious clients, FedZZ sustains an accuracy of 67.43%, while the accuracy of the second-best solution, FL-Defender, diminishes to 43.36%.

## 1 INTRODUCTION

Federated Learning (FL) is a decentralized machine learning (ML) paradigm [1] that facilitates model development using data from various sources while preserving their data privacy [5]. It is being increasingly adopted in areas with sensitive data, such as Google's Gboard [16] for next-word prediction, medical image recognition [31], etc.

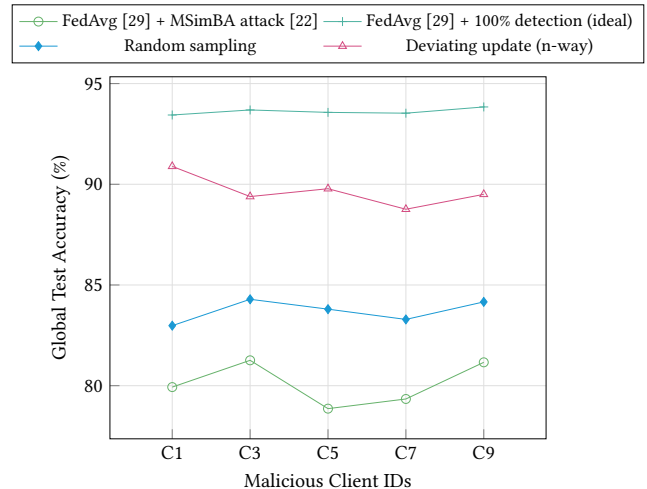

Figure 1: Performance of deviating update against data poisoning attack in FL.

**Poisoning FL system.** The data-opaque nature and offloaded training makes FL susceptible to attacks by malicious clients [11]. One or more adversarial clients can corrupt the jointly trained global model by sending malicious updates [21]. These attacks are further classified into (i) *Data poisoning:* Here, compromised clients poison training data without interfering with the local model training. (ii) *Model poisoning:* Here, compromised clients poison the model instead of data and provide malicious updates to the server. In this paper, we concentrate on *untargeted data poisoning attacks* within the realm of FL, as this aligns with typical FL production scenarios and constitutes a common threat [37]. Specifically, our focus is on *generic misclassification (untargeted)* rather than *specific misclassification (targeted) and under nobox attack settings*. Several works try to defend against data poisoning attacks. However, as we will explain in Section 2, most of them compromise privacy, suffer from overfitting, and/or require prior knowledge of poisoned samples. In this paper, we developed a robust defense against data poisoning attacks without the above drawbacks, outperforming all the state-of-the-art techniques. Furthermore, we comprehensively assess our approach under varying degrees of non-IID distributed data to gauge its performance in real-world data distribution settings.

**Deviating update hypothesis.** Our defense is based on the hypothesis that the updates from a malicious client doing data poisoning ($C_{dp}$) will differ from other benign clients' updates, and using $C_{dp}$ decreases the accuracy of the global model.

Based on the above hypothesis, we implemented a simple defense against a single client attack. Given $n$ clients, in each round, on the server side, we computed $n$ aggregations (one for each client), *i.e.*, $a_1, a_2,...,a_n$, where each $a_x$ is computed by aggregating updates from all clients except for $x$'s update. Second, a client update will be used for the global model *iff* it does not differ significantly in terms of cosine similarity from other updates. This process continues for each round, where we perform n-way aggregation of the updates, and a client's update will be used for the global model aggregation if it does not differ from the aggregated update from other clients. As shown in Figure 1, a preliminary experiment of this simple defense works well and outperforms existing solutions. But this simple solution has the following drawbacks. (i) *Scalability:* Performing n-way aggregation for every round does not scale for a large number of clients. (ii) *Handling multi-client poisoning attacks:* n-way aggregation does not handle cases with multiple malicious clients, as all aggregations omit only a single client update. For instance, if $C_1$ and $C_2$ are malicious, then we should have an aggregation without the updates $UP_1$ and $UP_2$ so that they can be detected as malicious and discarded. But this is not possible in n-way aggregation.

**Ideal solution.** We need a way to identify the group of benign clients whose aggregated update differs noticeably (with difference $\alpha$) from malicious client updates with data poisoning attacks. We call this the problem of finding a Poisoning Detecting Aggregation Group (PDAG). The aggregated update from PDAG can be used to defend against these attacks. As the clients can become malicious over time, the composition of PDAG might change over FL rounds. We acknowledge that finding PDAG, especially in active multi-client data poisoning attacks, is hard or rather intractable.

**Our approach: Zone Based Deviating Update (ZBDU).** Instead of finding and maintaining a single PDAG, our technique finds and maintains multiple potential PDAGs, which we call *Zones*. Our approach splits all $n$ clients into $m$ disjoint zones (*i.e.*, $z_1,..,z_m$), where each $z_x$ has $\lfloor \frac{n}{m} \rfloor$ clients, distributing them evenly across the zones. In each round, the server performs $m$ zone level aggregations ($AZ_1,..., AZ_m$) - one for each zone - using update from clients in the corresponding zone. We then compare each client update with the zone level aggregation (discriminator aggregation) of a zone that *does not* contain the client by using cosine similarity metric (*cosim*). For instance, update from client 1, *i.e.*, $UP_1$ will be compared with $AZ_x$, iff $C_1 \notin z_x$. Finally, we will use a client update for the global model aggregation at the server only if it does not differ significantly (within $\alpha$) from the discriminator aggregation, *i.e.*, $cosim(UP_1, AZ_x) > \alpha$. We use a precision guided approach to group clients into Zones. Specifically, given $m$, our precision guided approach will actively group the clients into $m$ zones, such that the global model built using ZBDU is robust against data-poisoning attacks by frequently discarding updates from malicious clients. In addition, our method serves as a robust client selection technique under highly non-IID data and no attack settings. Furthermore, we show that our technique also has the formal guarantee of monotonically increasing the accuracy of the global model. We implemented our approach in a tool called FEDZZ, which can be easily integrated into existing FL systems and has no measurable overhead. Our evaluation shows that FEDZZ is an effective and efficient mitigation method against data poisoning attacks by outperforming the existing state-of-the-art methods. In summary, the following are our contributions:

- We propose Zone Based Deviating Update (ZBDU), a new defense against untargeted data poisoning attacks.
- We design and implement FEDZZ, which uses a precision-guided approach to create and maintain client zones, an important requirement for ZBDU.
- We extensively evaluated FEDZZ with various attack configurations and show that FEDZZ effectively mitigates data poisoning attacks by maintaining high accuracy (∼67%) even when the attack rate is as high as 50%, outperforming the existing state-of-the-art techniques. Our comparative evaluation also shows superior detection rates for FEDZZ, with more than 80% detection rate (v/s 50% by existing techniques) and less than 45% false positive rate (v/s 80% for existing techniques) for varying attacks.
- We made our implementation available as open source at github.com/NaveenKumar-1311/FEDZZ to assist future research.

## 2 RELATED WORK

In this section, we will present a broader related work in the area of defenses in FL and the use of fuzzing in FL or in general Machine Learning (ML).

### 2.1 Existing Defenses and their Limitations.

Current defense strategies against FL data poisoning attacks typically fall into two categories: anomaly detection [38] or Byzantine robust model aggregation techniques [4, 42] that significantly reduce the impact of malicious updates. The anomaly detection approach involves classifying various elements of client updates and discarding updates that deviate from the dominant group. Techniques employed for this purpose vary, depending on the specific attributes of the client update group in question [40]. For instance, AUROR [38] employs clustering to identify indicative features, which are then used to filter out toxic updates. Sageflow [34] addresses both straggler clients and adversaries through methods such as staleness-aware grouping, entropy-based filtering, and loss-weighted averaging.

Recently, DeepSight [35], proposed by Rieger *et al.*,, underscores the significance of analyzing updates in neural networks. This method incorporates techniques like NormalizEd UPdate energies (NEUPs), which evaluate the total magnitude of parameter updates for the output layer. However, this also poses *a substantial privacy concern for client data. The output layer could potentially reveal information about the label frequency distribution in the training data, rendering it vulnerable to label inference attacks [10]*. In contrast, our approach leverages a trusted execution environment (TEE) [8, 30]. The client-side comparison of local and global model updates occurs within this environment, and a discard flag is communicated to the server. This way, the server is relieved of the responsibility of comparing updates, preventing the possibility of label inference attacks. Moreover, our defense methodology operates in a *black-box* manner, thereby negating the need for any information about the trained model or the learning task. This agnostic nature contributes to its widespread applicability and acceptance.

Few techniques like FLTrust [6] and SIREN [14], operate under the assumption that the service provider maintains a clean and small training dataset at the server, termed the "root dataset." This dataset should adhere to the constraint of being distributed similarly to the overall training data distribution across all clients. Additionally, the service provider establishes a "server model" based on this dataset to initiate trust. Furthermore, collecting a reliable root dataset can be a challenging task, and FLTrust's performance suffers when there is a significant deviation in the distribution of the root dataset compared to the training dataset [45]. On the contrary, our approach does not require any such root dataset on the server. Further, defenses based on novel aggregation techniques [4, 32] require significant changes to the underlying learning algorithm used in FL systems resulting in substantial deployment overhead. As shown by the recent work [37], a practical (and applicable to real systems) defense technique against untargeted data poisoning attacks that do not affect the privacy of clients' data should be agnostic to the learning technique (*i.e.*, black-box) used in a FL system so that the defense is applicable to all FL systems. This requirement forces us to develop an anomaly-based technique as it can be agnostic to the learning algorithm used in FL.

## 2.2 Fuzzing in Deep Learning

Fuzzing or fuzz testing [12, 27, 28] is an automated software testing technique widely used to find system program failures. Fuzzing is proved to be very effective in identifying vulnerabilities and threats in software systems [7, 15, 19]. Few works, such as TensorFuzz [33] and DeepHunter [41], have tried to use fuzzing to detect potential defects of general-purpose deep neural networks. However, to our knowledge, there is no explicit use of fuzzing in FL to defend against data-poisoning attacks. We map the fuzzer-generated inputs to zones and use the precision of the learned model to guide the zone selection. The use of fuzzing to generate a sequence of tokens has been explored before to fuzz interpreters [36]. However, instead of modifying the input generation, we perform post-processing of the generated input to map clients to different zones. Our approach allows the fuzzer to fully use its input generation ability and enables FEDZZ to be easily configurable to use other fuzzers.

## 2.3 Guided Mutation for Input Generation

Mutation techniques, such as crossover and bitflips [43], are commonly used as an effective way to generate interesting inputs, especially in the domain of automated software testing (*e.g.*, Fuzzing [27]).

An optimization usually guides the use of mutation techniques. For instance, in the case of fuzzing, the optimization is usually code coverage [39]. Consequently, the fuzzer uses mutation techniques on existing or base input that can potentially improve code coverage. Similar to FL, the mutation-based input generation is an iterative process, and the selection of mutation strategies changes over time based on their effectiveness [25].

## 3 THREAT MODEL

We derive our threat model from realistic FL deployments (*e.g.*, Google's Mobile keyboard Prediction [16, 37]), where clients provide the data for local model training. The clients *cannot interfere with the training procedure i.e., trusted execution environment (TEE) [8, 30]*, and the communication with the server occurs *through an encryption channel and hence cannot be interfered with*. As shown by the recent work [21, 37], *our threat model is the most realistic and of practical use in FL*. Based on this, we present the goals and capabilities of the attacker and the assumptions we make about the FL setup.

*3.0.1 Attacker Goal.* The main goal of the attacker is to make the global model (*i.e.*, the one used to perform testing on the server) mispredict/misclassify data and thereby have a reduced global test accuracy (GTA). The attacker is interested in *generic misclassification (untargeted) rather than specific misclassification (targeted)*.

*3.0.2 Attacker Capabilities.* We assume the attacker has the following capabilities on the server and compromised clients.
**Server side.** We assume the server is trustworthy, incurious about updates, and a black-box to the attacker. As such, the attacker has *no access to parameters or predictions of the global model* as we focus on training time (causative) attack.
**Client side.** We assume that the attacker controls the data used in one (single-client attack) or more clients (multi-client attack). Precisely, the attacker can only manipulate the local data of the compromised clients with no access to the compromised clients' training procedure or communication with the server. However, the attacker can access the predictions of a compromised client's local model on any chosen data. We assume an active attacker with a repeat and adaptive poisoning of the compromised clients' data, so the attack persists over the entire FL training. In summary, the attacker *has control of all the data provided to train a local model on compromised clients and can also know the predictions of these clients' local model on any chosen data*. But the attacker can *neither interfere with the local model's training process nor access its internal parameters*. Clients' local training mechanism communicates with the server over an encrypted channel and hence cannot be interfered with. Figure 2 shows the client side execution in FEDZZ.

Using the terms defined in the recent work [37], our threat model can be called *nobox* (no access to server side), *online* (can modify client side data continuously), data poisoning attack. The goal of FEDZZ is to maximize the global test accuracy by identifying client groups to aggregate by dropping the updates from malicious clients.

## 4 PROPOSED METHOD

FedZZ works by making a few modifications to the existing FL system's workflow. Algorithm 1 shows our FedZZ integrated FL system. We present a summary of our adopted notations in Table 3 in the Appendix. We start by describing the modifications needed to implement ZBDU and then our precision guided approach to identify the necessary parameters by using guided mutations for input generation (Section 2.3). We split our modifications into those on the server side and the client side. We use $C$ to represent the set of all clients, where $n$ indicates the total number of clients.

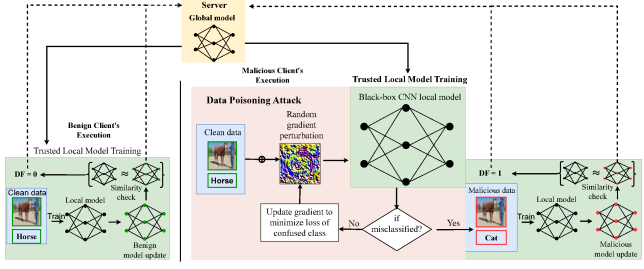

**Figure 2: Client side execution in FedZZ.** The local training on clients happens in a trusted and isolated container. After training, the client performs similarity check of model update and sends $DF$ flag to the server in each round (or epoch). Malicious clients performs active data poisoning attack by using the predictions of the local model and adding gradient noise (MSimBA [22]).

### 4.1 Server side

The server maintains the *Zones* set ($Z$) and the *discriminator zone map* ($\Psi$). Where $Z$ is a set of $m$ disjoint client groups of the same size called *Zones*. Formally, $Z = \{z_1, z_2, ..., z_m\}$, such that $\forall i, j \in [1, m], ((|z_i| = |z_j|) \land ((z_i \cap z_j) = \emptyset)$ and all clients belong to a zone i.e., $\forall i \in [1, n], \exists j \in [1, m] \mid c_i \in z_j$. For instance, for nine ($n = 9$) clients, $Z = \{z_1 = (c_1, c_3, c_4), z_2 = (c_5, c_2, c_8), z_3 = (c_6, c_7, c_9)\}$ represents a valid zone set, whereas $Z = \{z_1 = (c_5, c_3, c_4), z_2 = (c_1, c_2, c_8), z_3 = (c_6, c_3, c_9)\}$ is *invalid* as $c_3$ is present in two zones i.e., $z_1$ and $z_3$ and also $c_7$ does not belong to any zone. The discriminator zone map, $\Psi$, is a map from a client to its discriminator zone, i.e., $\Psi : C \to Z$. Every client has an entry in $\Psi$, formally $\forall i \in [1, n], c_i \in \Psi$. Furthermore, for a given $Z$, $\Psi$ always maps a client to a zone that *does not* contain the client, formally, $\forall i \in [1, n], (\Psi(c_i) == z_j \iff c_i \notin z_j)$. For the given valid zone above, $\Psi(c_1) = z_2$ is valid, where as $\Psi(c_1) = z_1$ is *invalid* because $c_1 \in z_1$.

At each epoch, from each client $i$, the server receives ($UP_i, DF_i$), where $UP_i$ is the client side update as in regular FL systems and $DF_i$, is the *discard flag*, a boolean value which indicates whether the update should be discarded (`true`) or not (`false`). The server aggregates all the client updates whose discard flag is not set and uses it to update the global model. Specifically, at each epoch, we use synchronous federated weighted average aggregation [29] to compute the new parameters for the global model ($w^G_{new}$) as, $w^G_{new} = \sum_{k \in [1,n]} \lambda_k UP_k \mid DF_k == 0$, where $\lambda_k = \frac{len(\mathbb{D}_k)}{\sum_{k \in [1,n]} len(\mathbb{D}_k)}$), and $\sum_{k \in [1,n]} \lambda_k = 1$. Next, the server uses $Z$ to compute $m$ zone

**Algorithm 1** Standard FL with **our** FedZZ **framework**

1: **Input:** Global model $w^G$, local client data $\mathbb{D}$, zones set $Z$, discriminator zone map $\Psi$, learning rate $\eta$, loss function $l$
2: **Output:** Global test accuracy $GTA$
3: $Z \leftarrow$ random zones set
4: **Client execution** ($AZ$):
5: **for** each client $i = 1$ **to** $n$ **do**
6:    $UP_i = LM_i - \eta \Delta l(AZ_i, \mathbb{D}_i)$
7:    $cosim_i \leftarrow \frac{UP_i \cdot AZ_i}{||UP_i|| \cdot ||AZ_i||}$
8:    **if** $cosim_i < \alpha$ **then**
9:      $DF_i = 1$
10:   **else**
11:     $DF_i = 0$
12:   **end if**
13:   **return** $UP_i, DF_i$
14: **end for**
15: **Server execution** ($UP, DF$):
16: $w^G_{new} = \sum_{i \in [1,n]} \lambda_i UP_i \mid DF_i == 0$
17: Compute $GTA \leftarrow$ Test ($w^G_{new}, \mathbb{D}_{test}$)
18: $Z_{new} \leftarrow$ Zones calibrator ($Z, GTA$) (Refer **Algorithm 2**)
19: **for** each zone $j = 1$ **to** $m$ **do**
20:   $AZ_j = \{\sum_{i \in [1,n]} \lambda_i UP_i \mid \forall c_i \in z_j\}; z_j \in [Z^1_{new}, Z^m_{new}]$
21: **end for**
22: **for** each client $i = 1$ **to** $n$ **do**
23:   **for** each zone $j = 1$ **to** $m$ **do**
24:     **if** $\Psi(c_i) = z_j$ **then**
25:       **return** $AZ_j$
26:     **end if**
27:   **end for**
28: **end for**
29: **return** $GTA$

level aggregations, i.e., $\{AZ_1, ..., AZ_m\}$, by aggregating the updates from clients that belong to the corresponding zone (irrespective of their discard flag). Finally, these zone level aggregations will be sent to the clients based on their mapping in the discriminator zone map, $\Psi$. Specifically, for a client $i$ if $\Psi(c_i) = z_j$, then $AZ_j$ will be sent to the client, as shown in Algorithm 1.

*4.1.1 Configuring $Z$ and $\Psi$.* The crux of FedZZ lies in configuring server with effective *Zones* set ($Z$) and *discriminator zone map* ($\Psi$). We configure $\Psi$ to have a fixed adjacent zone mapping, *i.e.,* we map every client $i$ to the zone next to the one it belongs to. Formally, $\forall i \in [1, n] \mid (c_i \in z_j) \implies (\Psi(c_i) == z_{(j+1)\%m})$. For instance, for nine clients ($n = 9$) with three zones ($m = 3$), if client $c_1$ belongs to $z_2$ then $\Psi(c_1) = z_3$. Similarly, if $c_3$ belong to $z_3$ then $\Psi(c_3) = z_1$. Further, to tackle the accuracy drop issue in the discriminator zone map ($\Psi$), especially when neighbouring zones involve malicious clients with comparable updates that could lead to identification issues, we've developed a precision-oriented approach, detailed in the upcoming sections. This method adeptly deals with the problem of nearby malicious clients within the discriminator zone map. Our approach effectively detects malicious behaviour by gradually adjusting and including effective zone set configurations. This enhancement significantly strengthens our defense against data poisoning attacks in FL.

**Challenge.** Finding an effective $Z$ for ZBDU that can help in thwarting data poisoning attacks is a hard problem. For a given $n$ and $m$, the number of possible zone maps is given by $\frac{n!}{((n/m)!)^m}$. Even for $n = 9$ and $m = 3$, there are 1,680 possible zone maps. In each epoch, all clients should train their local model and send the corresponding update to the server for all possible zone maps. This is impractical in a real-world setting where we have a large number of clients, and local training on the client side takes a considerable amount of time and resources. Hence, we propose a precision guided technique to identify $Z$, such that the accuracy of the resulting global model built using the aforementioned ZBDU technique is maximized even in the presence of data poisoning attacks. We use an iterative and continuous approach to configure $Z$ during FL to be resilient to an active adversary who continuously performs data poisoning attack in FL.

*4.1.2 FedZZ Zones Calibrator.* The server will invoke the zones calibrator after every $\xi$ epochs to get a new zones map ($Z_{new}$). We start with a random initialization of $Z$. The zones calibrator maintains a priority queue of interesting inputs (*IInputs*), with a list of zone maps ordered according to the decreasing order of global test accuracy (GTA). The Algorithm 2 shows the pseudo-code of our zones calibrator. The zones calibrator runs for $\tau$ iterations and starts by adding the current zone set ($Z$) to *IInputs*. In each iteration, we run mutation (`mutate`) using *IInputs* to get $Z_{mut}$. As shown in Figure 3, after every $\xi$ epochs, the server invokes FedZZ zones calibrator with the current configuration of $Z$. Here, the probability of picking an input (*i.e.*, zone set) to mutate is proportional to its order in the list. We invoke the server with $Z_{mut}$, which will be used to mimic FL using the client updates ($DF = 0$), and the resulting $GTA_{mut}$ is returned. If $GTA_{mut}$ is greater than the last known best ($GTA_{max}$), we will save the $Z_{mut}$ as the best zones set $Z_{new}$. The process continues for $\tau$ iterations, and the best zones set ($Z_{new}$) will be returned. However, as we show in Section 6.3, occasionally discarding updates from benign clients does not significantly affect the global model test accuracy. Over time, our technique learns to detect the malicious clients. We integrated the FedZZ zones calibrator by making modifications to the AFL++ implementation [43].

Further, FedZZ is independent of the type of learning (task) employed inside the client training procedure. Our ZBDU approach is specifically designed to be agnostic to any internal learning mechanism. It focuses solely on the similarity between updates in raising the discard flag on the client side. This makes FedZZ a versatile and adaptable approach that can be easily extended to any computer vision or NLP tasks, regardless of the underlying learning technique.

## 4.2 Client side.

As in general FL systems, every client uses the update sent by the server (*i.e.*, $AZ_i$) to train its local model using the local data. Clients compute the update (*i.e.*, $UP_i$) made to the local model and compare this update with the server update, *i.e.*, $AZ_i$. The $DF$ flag will be set if the new update differs by more than a pre-configured deviation threshold ($\alpha$). It can be given as $DF = 1, if (cosim(UP_x, AZ_x) < \alpha)$, indicative of potential deviations of malicious updates. Here, *cosim* is the cosine similarity. The utilization of cosine similarity as a pivotal metric within the proposed approach is rooted in its inherent

**Algorithm 2** Zones Calibrator

1: **Input:** Current Zones set ($Z$), Global test accuracy ($GTA$)
2: **Output:** New Zones set ($Z_{new}$)
3:
4:    // Priority queue of inputs
5:    $IInputs \leftarrow IInputs \cup (Z, GTA)$
6:    // Best accuracy and zones map
7:    $GTA_{max} \leftarrow GTA$
8:    $Z_{new} \leftarrow Z$
9:    **for** $i = 1$ **to** $\tau$ **do**
10:       $Z_{mut} \leftarrow$ `mutate`($IInputs$)
11:       $GTA_{mut} \leftarrow$ `Server`($Z_{mut}$)
12:       $IInputs \leftarrow IInputs \cup (Z_{mut}, GTA_{mut})$
13:
14:       // Update the best zone set
15:       **if** $GTA_{mut} > GTA_{max}$ **then**
16:          $GTA_{max} \leftarrow GTA_{mut}$
17:          $Z_{new} \leftarrow Z_{mut}$
18:       **end if**
19:    **end for**
20:    **return** $Z_{new}$

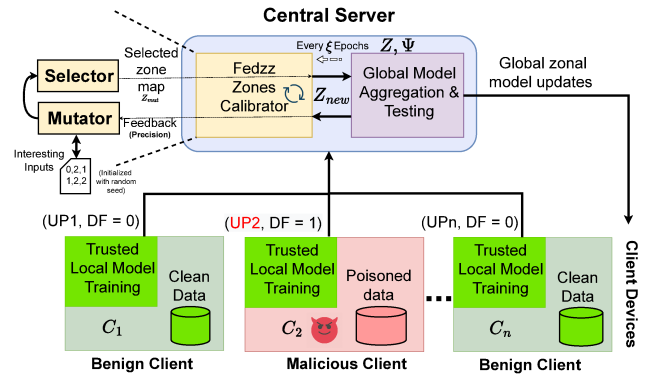

**Figure 3: Overview of FedZZ integrated into a FL system with $n$ clients ($C_1, C_2, \ldots, C_n$). The malicious client ($C_2$) poisons the training data. The trusted local model training checks for malicious update and raises discard flag ($DF$) as 1 based on the received zone level update. The central server receives the model updates along with the discard flag and performs aggregation based on the discard flag. After every $\xi$ epochs, the zones calibrator will be called to compute a new zone map.**

strengths that resonate effectively with the task of identifying malicious clients within the context of FL scenarios [2, 11, 24, 26, 44]. Its efficacy is derived from its ability to assess direction alignment between vectors, independent of their magnitudes. This characteristic has been harnessed in various defense schemes to detect adversarial activities within FL environments. A pertinent example lies in the work of Fung *et al.* [11], who introduced the FoolsGold defense mechanism against sybil attacks by evaluating cosine similarity between aggregations of clients' historical updates. Malicious updates exhibit significantly higher similarities compared to benign ones, thus establishing a discernible distinction. Similarly, Bagdasaryan et al. [2] employ cosine similarity to gauge the congruence between the local model and the global model, utilizing a boosting factor to amplify the similarity for malicious updates, enabling their detection. Furthermore, the defense strategies presented by Lu et al. [24] encompass both model pre-aggregation with cosine similarity assessment and backdoor neuron activation. Notably, the pre-aggregated model closely approximates the backdoor model upon successful attacks, resulting in elevated cosine similarity, a value that tends to deviate substantially for benign updates.

Aligned with these research insights, our work leverages the cosine similarity metric to compare updates from locally trained and global models robustly, ultimately generating the discard flag. The strength of this approach lies in the contextual constraints of the threat model, wherein the client's capability is confined to providing data without the ability to influence the training procedure. This fortifies the defense mechanism, rendering it resilient against adaptive data poisoning attacks. The inherent dynamics of successful attacks become discernible by juxtaposing the locally poisoned model, crafted through data poisoning attack, with the global aggregated model encompassing all other updates. Typically, the divergence in directions renders benign updates markedly distinct, consequently resulting in smaller cosine similarity. *In summary, the employment of cosine similarity within this defense framework is substantiated by its proven efficacy in discerning malicious clients and deviations caused by data poisoning attacks in FL paradigms, as demonstrated by prior research endeavours [2, 11, 24, 26, 44].* Finally, the new update, $UP_i$, and the $DF$ flag will be returned to the server, as shown in Algorithm 1. As mentioned in Section 3, the client can only provide data and cannot affect the training procedure.

Figure 2 shows the client side execution. In the case of benign clients, the local update will not differ much from the zone level aggregated update, and consequently, $DF$ will not be set - which will cause the server to use the update for aggregating into the global model. Whereas in the case of malicious clients using poisoned data, the local model update differs noticeably and sets the $DF$, causing the server to discard the update.

### 4.3 Convergence Analysis of FEDZZ

The theorem provides mathematical proof for the convergence of the FEDZZ zones calibrator towards the local minimum of the loss function.

THEOREM 4.1. *Let $\mathcal{L}$ be the global cross-entropy loss function in relation to the zone map generated by the FEDZZ zones calibrator using the AFL++ algorithm [43]. The sequence of cross-entropy loss values of the global model $w^G$ obtained from the generated zone map $Z$ at a global epoch $t$, denoted as $\mathcal{L}(w_t^G \mid Z_t)_{t \geq 1}$, is non-increasing and will converge to the local minimum $\mathcal{L}(w^{G^*} \mid Z^*)$ in finite time.*

PROOF. We use mathematical induction to prove the theorem.

**Basis step:** When $t = 1$, FEDZZ zones calibrator using the AFL++ algorithm generates the initial zone map $Z_0$, which is a random mapping. Since the cross-entropy loss function is non-negative, we have $\mathcal{L}(w_1^G \mid Z_1) \leq \mathcal{L}(w_0^G \mid Z_0)$, and the sequence is non-increasing. Also, the AFL++ uses a genetic algorithm to explore the state space and generate zone maps that result in a higher global test accuracy (feedback to AFL++) and a lower cross-entropy loss. Therefore, we have $\mathcal{L}(w_1^G \mid Z_1) \leq \mathcal{L}(w_0^G \mid Z_0)$.

**Induction step:** Assume that for some $k \geq 1$, the sequence $\mathcal{L}(w_t^G \mid Z_t)_{t \geq 1}^T$ is non-increasing, i.e., $\mathcal{L}(w_1^G \mid Z_1) \leq \mathcal{L}(w_0^G \mid Z_0), \ldots \mathcal{L}(w_T^G \mid Z_T) \leq \mathcal{L}(w_{T-1}^G \mid Z_{T-1})$ and converges to the local minimum $\mathcal{L}(w^{G^*} \mid Z^*)$. We need to show that, the statement holds true for $t = T + 1$, i.e., $\mathcal{L}(w_{T+1}^G \mid Z_{T+1}) \leq \mathcal{L}(w_T^G \mid Z_T)$.

Since $\mathcal{L}$ is a continuous function of $w_G$, and $Z$ is a zone map generated by the AFL++ algorithm, there exists a small enough positive constant $\theta$ such that $\mathcal{L}(w_{T+1}^G \mid Z_T + \theta(Z_T - Z^*)) \leq \mathcal{L}(w_T^G \mid Z_T)$. This means that the AFL++ algorithm can generate a zone map

$$Z_{T+1} = Z_T + \theta(Z_T - Z^*) \quad (1)$$

such that $\mathcal{L}(w_{T+1}^G \mid Z_{T+1}) \leq \mathcal{L}(w_T^G \mid Z_T)$. Since the goal of AFL++ is to find a zone map that minimizes $\mathcal{L}$, which corresponds to finding a zone map that performs well in discarding malicious clients, at each calibration round. Hence, we can find a new zone map $Z_{T+1}$ that has a lower value of $\mathcal{L}$ than the previous zone map $Z_T$.

In particular, the Eq. 1 implies that we can always find a small enough perturbation $\theta$ to the previous zone map $Z_T$ such that the new zone map $Z_{T+1} = Z_T + \theta(Z_T - Z^*)$ (where $Z^*$ is the optimal zone map) has a lower value of $\mathcal{L}(w^G)$ than the previous zone map $Z_T$. This perturbation is chosen in the direction of $Z^*$ (*i.e.*, towards the optimal zone map), which ensures that the algorithm is making progress towards the optimal zone map. Therefore, we have $\mathcal{L}(w_{T+1}^G \mid Z_{T+1}) \leq \mathcal{L}(w_T^G \mid Z_T)$, as the sequence is non-increasing. By the induction hypothesis, it is clear that the sequence $\mathcal{L}(w_t^G \mid Z_t)_{t \geq 1}^T$ is non-increasing and converges to the local minimum $\mathcal{L}(w^{G^*} \mid Z^*)$.

Finally, by mathematical induction, we conclude that the sequence $\mathcal{L}(w_t^G \mid Z_t)_{t \geq 1}^T$ is non-increasing and converges to the local minimum $\mathcal{L}(w^{G^*} \mid Z^*)$, and the AFL++ algorithm converges to the local minimum of the cross-entropy loss function. Since the cross-entropy loss is a non-negative function, this sequence must converge to a finite value. Thus, the AFL++ algorithm guarantees that the system will converge to the desired state $Z^*$ in finite time. In addition, as there are a finite number of possible zone maps $\frac{n!}{(n/m)!^m}$, the AFL++ algorithm generates zone maps that result in a decreasing cross-entropy loss over time and guarantees that the system will reach the desired state $Z^*$ in finite time. □

## 5 IMPLEMENTATION

We implemented FEDZZ zones calibrator by modifying AFL++. This simplified our implementation, as AFL++ already has various mutation strategies and uses techniques, such as fork server, to achieve very high execution rates. The FL server is modified to expose a RESTful interface that encapsulates the computation of the feedback for a given $Z_{mut}$. Our modified AFL++ communicates with the REST interface to get the feedback (*i.e.*, precision in the form of global test accuracy) for each mutated input (*i.e.*, $Z_{mut}$). This separation of FL server execution, in addition to providing modularity, also enables users to configure the feedback metric or develop a custom metric without worrying about the internals of FL execution. Further, the entire mutator selector setup is unaware of the type of machine learning model, dataset, and execution and treats the server as a black-box.

Table 1: Experimental details

| Name | Value |
| --- | --- |
| Clients | 40 |
| Classification model | ResNet18 [17] |
| Global epochs | 300 |
| Local epochs | 5 |
| non-IID parameter | 0.1, 0.5, 1 (default), 5, 10 |
| Batch size | 64 |
| $\eta$ | 0.01 |
| $\alpha$ | 0.80, 0.90, 0.95, 0.97 (default), 1 |
| Attack methods | MSimBA [22], DPA-SLF [37], and DPA-DLF [37] |
| Attack settings | single-client, multi-client |
| Single-client attack IDs | $C_1$, $C_3$, $C_5$, $C_7$, and $C_9$ |
| Multi-client attack configuration | 20%, 30%, 40%, and 50% malicious clients |

## 6 EVALUATION

We use the following research questions to guide our evaluation.

- **RQ1- Effectiveness:** How effective is FedZZ in mitigating data poisoning attacks by building a precise global model?
- **RQ2- Comparison with the State-of-the-art:** How effective is FedZZ compared to existing state-of-the-art approaches?
- **RQ3- Ability to perform under varying degree of non-IID:** How effective is FedZZ compared to existing state-of-the-art approaches with malicious updates and non-IID benign updates?
- **RQ4- Detection Rate:** How effective is our ZBDU technique in identifying clients performing data poisoning attacks?
- **RQ5- Overhead:** What is the overhead of using FedZZ in existing FL systems?

### 6.1 Datasets and FL Setup

We utilize two standard classification datasets for our evaluation, namely, CIFAR10 [20] and EMNIST [9]. CIFAR10 comprises 60,000 samples with ten classes. EMNIST consists of 671,585 samples of handwritten characters and digits distributed across 62 classes, encompassing upper and lowercase handwritten characters. We partition these datasets into 80% train and 20% test sets. The server test set is employed to calculate the global model accuracy. To maintain a consistent image size, we scale the images in these datasets to an average resolution of 224 × 224. Table 1 presents the details about the FL setup and other parameters. We conducted each experiment five times and presented the results and graphs as the average outcomes of these five simulations. We used Python version 3.6 with frameworks like PyTorch, pandas, and NumPy. We implemented the experiments such that the local model training at the clients and global model testing at the server happens on the Nvidia Tesla M60 GPU with 8GB RAM.

### 6.2 Baselines and FedZZ Configurations

We use the following baselines and configurations of FedZZ to evaluate its effectiveness:

- *FedAvg [29]:* FL with FedAvg aggregation and no defense. Ideally, FedZZ should perform similar to this baseline under **no attack** scenarios.

- *FL100 (Ideal):* FL system with max possible accuracy. Here the server knows the malicious client ids, and the detection rate is 100%. This represents the *upper bound for defense techniques*.
- *Random Sampling (RS) of the Clients:* This represents FL system with random sampling of clients for every round. As FedZZ involves generating zones and dropping the malicious client updates. It should perform better than the server's random sampling of the clients at every communication round.
- *n-way:* This represents our hypothesis of n-way aggregation (Section 1), where we always have $n$ zones, where each zone contains clients except one. It is the expected performance of FedZZ in single client attacks. However, we expect this to perform poorly in multi-client attack cases.
- *FedZZ Configurations:* We use three configurations of FedZZ with different number of zones. Specifically, FedZZ10 ($m = 10$), FedZZ5 ($m = 5$), and FedZZ2 ($m = 2$).

### 6.3 Effectiveness

We use the Global Test Accuracy (GTA) as the metric to evaluate the effectiveness of our approach (FedZZ10, FedZZ5, FedZZ2) against other configurations. Specifically, we compute $GTA$ as the accuracy of the global model after stabilization (*i.e.,* 300 global epochs) on $\mathbb{D}_{Test}$. Formally, $GTA = \frac{TP+TN}{|\mathbb{D}_{Test}|} \times 100$, where $TP$, $TN$ are true positives and true negatives given by the aggregated global model. For a robust defense technique, $GTA$ is expected to be similar to FL100's $GTA$ (the ideal defense) under different attack settings.

**No Attack or Base Case:** We tested FedZZ configurations with no attack scenario and observed that *the GTA is mostly the same, in fact, higher in a few cases than the base FL*. This is expected because, as shown by the recent work [3], we often do not need updates from all clients to build a precise model in FL.

**Attack Cases:** Figure 4 and Figure 5 present the results of different techniques under single and multi-client attack settings, respectively.

*Single-client attack:* In the CIFAR10 dataset (Figure 4a, 4b), first, FedZZ configurations outperform random sampling demonstrating the base effectiveness of our technique. As expected (Section 1), n-way aggregation performs best and is close to the ideal defense (*i.e.,* FL100). It is interesting to see that the effectiveness of FedZZ10 and FedZZ5 are relatively the same, with FedZZ5 being superior in a few cases. However, the effectiveness drops as we further decrease the number of zones to 2 (*i.e.,* FedZZ2). As we decrease the number of zones, the effect of a client update on zone aggregation decreases. In the case of FedZZ10 with ten zones and each with four clients, a single malicious client can be in any of the ten zones and has a 1/4 (25%) contribution to the aggregation. For FedZZ5 with five zones and each with eight clients, a client has 1/8 (12.5%) contribution. Similarly, for FedZZ2, the contribution is further decreased to 1/20 (5%). However, the number of clients that receives a zonal update increases, which increases the probability of a client receiving an update from a zone. In the case of FedZZ10, a client receiving the update from a zone is 10% (4/40). For FedZZ5, its 20% (8/40) and for FedZZ2, its 50% (20/40).

In summary, as we *decrease the number of zones, both true positives and false negatives for detecting malicious clients increases*. Similarly, *increasing the number of zones increases false positives*

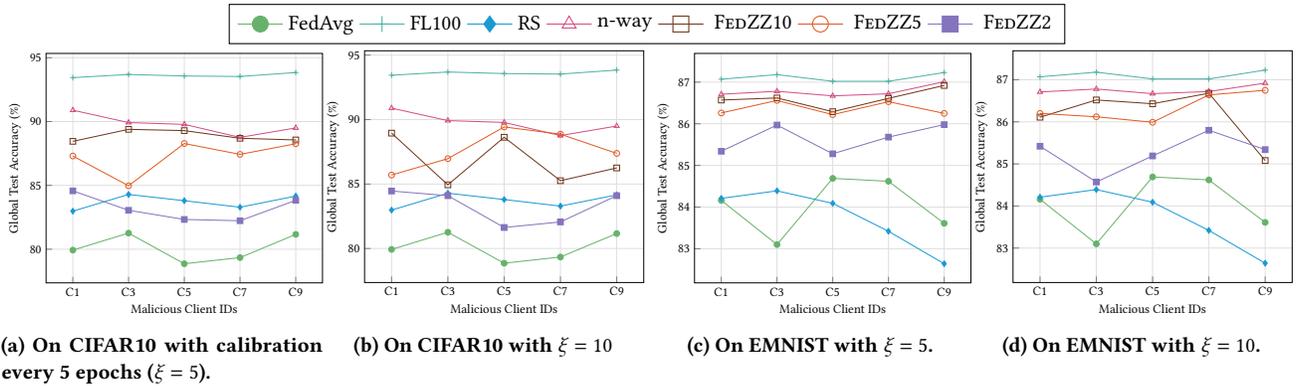

(a) On CIFAR10 with calibration every 5 epochs ($\xi = 5$).

(b) On CIFAR10 with $\xi = 10$.

(c) On EMNIST with $\xi = 5$.

(d) On EMNIST with $\xi = 10$.

Figure 4: *Effectiveness on single-client attacks*: Comparison of Global Test Accuracy ($GTA$) of various baselines across different datasets and calibration granularities.

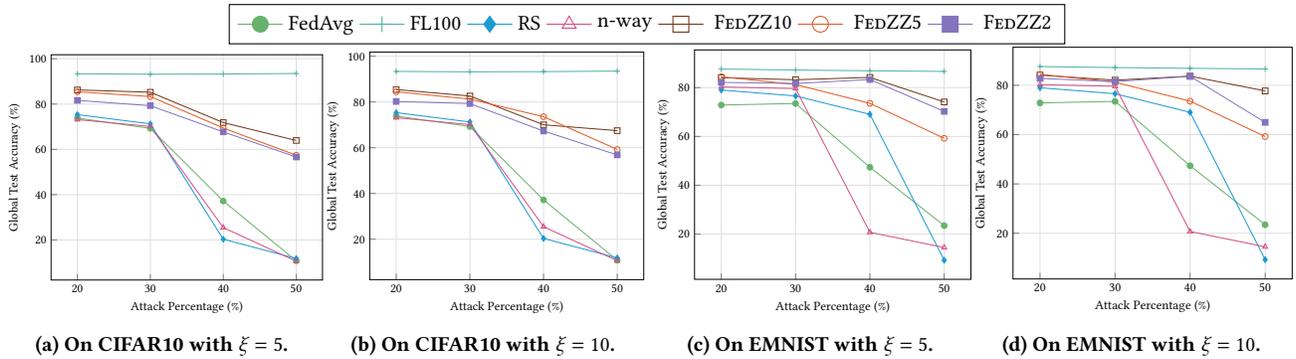

(a) On CIFAR10 with $\xi = 5$.

(b) On CIFAR10 with $\xi = 10$.

(c) On EMNIST with $\xi = 5$.

(d) On EMNIST with $\xi = 10$.

Figure 5: *Effectiveness on multi-client attacks*: Comparison of Global Test Accuracy ($GTA$) of various baselines across different datasets and calibration granularities.

*and true negatives.* Configuring the number of zones appropriately is important for the best performance. The trend is similar but more pronounced in the EMNIST dataset. Another interesting observation is that as we increase the calibration interval (*i.e.*, $\xi$), the effectiveness of FedZZ configurations decreases (Figure 4a v/s 4b and Figure 4c v/s 4d). This is expected because there are fewer opportunities to calibrate (60 (300/5) v/s 30 (300/10)). This allows the malicious clients to have more effect on the global model (*i.e.*, 5 v/s 10).

*Multi-client attack (Figure 5):* We observe that FedZZ continuously outperforms n-way and random sampling in multi-client attack settings as shown in Figure 5. As mentioned in Section 1, this is expected as the n-way of aggregating model updates results in $n$ aggregated models with multiple malicious updates. The FedZZ displayed robustness and was able to retain a high accuracy (∼ 77%) even under hard attack (50%) scenarios, compared to n-way and random sampling. The Figure 5a and 5b present the results on CIFAR10 dataset which show FedZZ10, FedZZ5, and FedZZ2 perform better than random sampling. Similar to the single client attack case, FedZZ5 has shown better performance compared to FedZZ10 and FedZZ2. The trend is more pronounced in the EMNIST dataset as shown in Figure 5c and 5d. We believe that the inter & intra-class variability of EMNIST and CIFAR10 datasets, along with a difference in the number of classes, can be the reason for this.

> **Answer to RQ1:** *In summary, as expected, n-way showed better $GTA$ in the single-client attack setting, and FedZZ is close to n-way with a difference of ≈ 2%. On the contrary, FedZZ outperformed n-way and other methods in the multi-client attack setting by a large margin.*

### 6.4 Comparison with Existing Defenses

We have selected the following state-of-the-art techniques based on their up-to-dateness and relevance. Then, we categorized them into four groups, including other baselines and configurations from Section 6.2. These categories are (1) 100% detection rate (FL100), (2) client selection techniques (RS, DivFL [3]), (3) aggregation techniques including byzantine robust (FedAvg [29], Krum [4], Trimmed mean (TM)[42], Median [42], FLTrust [6]), and (4) very recent defense methods including those from 2022 (FLAME [32]) and 2023 (LoMar [23], FL-Defender [18]). We have categorized these techniques to provide a comprehensive overview of the current state-of-the-art techniques in this field. Table 2 shows our comparison results, where we report the *GTA* of the global model.

*CIFAR10 dataset:* DivFL, with its submodular optimization, is able to outperform all other methods under no attack settings, whereas FedZZ is close with ($\leq 10\%$). However, under high non-IID non attack settings (as explained further in Section 6.5 and Appendix A.1), FedZZ has demonstrated its effectiveness as a strong client selection technique and outperformed DivFL. According to the findings in FLTrust [6], the performance is contingent on the similarity between the root dataset distribution and the overall training data distribution. Specifically, FLTrust performs well when the two distributions are not significantly divergent. However, when evaluated against other baselines with a given root data distribution at the server, FLTrust performed poorly for all attack settings. Under single-client attack, n-way is expected to perform better, but FedZZ is able to stand second compared to other defense methods with $\leq 2\%$. Furthermore, FedZZ *outperformed all the methods under multi-client attack settings*. The effectiveness of FedZZ is clearly observed with ($\geq 20\%$) improvement compared to existing defenses under higher attack percentage settings (40% and 50%). This shows the robustness of FedZZ in retaining the accuracy under hard attack settings.

*EMNIST dataset:* Although, under single-client attack, trimmed mean (TM) outperformed all methods, FedZZ is close to $\leq 1\%$. On the other hand, for the multi-client attacks with small attack percentages, 20% and 30%, LoMar can outperform other methods. However, FedZZ is close with $\leq 1\%$. The effectiveness of FedZZ can be observed at higher attack percentages of 40% and 50%, where other techniques, such as TM, is highly ineffective and result in very low accuracy of 22.42% and 6.71%. This is expected, as trimming model updates across all clients results in considering all malicious updates for aggregation. Hence, the aggregated global model performs poorly under higher attack percentages. We can see a similar trend with FLAME, LoMar, and other defense methods whose accuracy also dropped with large attack percentages. We observe that measuring the relative distribution of model updates is inadequate for approximating the nearest neighbour search in LoMar, resulting in poor scoring of model updates. This is primarily due to the high malicious rate of model updates, leading to a significant drop in accuracy. This shows that FedZZ can effectively detect the poisonous updates and helps in maintaining robust accuracy even under high attack percentages ($> 30\%$).

Nevertheless, LoMar and FL-Defender are comparable to our performance in a few cases, they share certain limitations concerning privacy and effective update differentiation. LoMar method includes the potential for privacy breaches due to analyzing update distributions. Moreover, LoMar may struggle with distinguishing between malicious updates and highly non-IID benign updates, which are prevalent in real-world FL scenarios. FL-Defender involves potential privacy concerns due to analyzing gradients. Additionally, FL-Defender's reliance on last-layer gradients might lead to challenges in distinguishing between sophisticated attacks and naturally occurring deviations in benign gradients. We also performed a comparative evaluation on other data-poisoning attacks

**Table 2: Comparison of global test accuracy (%) with existing defenses from different categories under MSimBA [22] attack. Bold result indicates the best result for setting.**

| Defense | CIFAR10 | | | | | | EMNIST | | | | | |
|---|---|---|---|---|---|---|---|---|---|---|---|---|
| | No Attack | 1A | 20% | 30% | 40% | 50% | No Attack | 1A | 20% | 30% | 40% | 50% |
| FL100 | 94.54 | 93.44 | 93.36 | 93.18 | 93.27 | 94.48 | 87.96 | 87.07 | 87.61 | 87.2 | 86.89 | 86.63 |
| RS | 94.57 | 82.98 | 75.33 | 71.26 | 20.33 | 11.81 | 87.41 | 84.21 | 79.01 | 76.58 | 69.05 | 9.29 |
| DivFL [3] | **94.86** | 74.12 | 72.08 | 71.19 | 43.14 | 11.17 | 87.26 | 86.05 | 84.32 | 83.64 | 67.16 | 38.01 |
| FedAvg [29] | 94.54 | 79.93 | 73.83 | 69.26 | 37.18 | 10.78 | **87.96** | 84.16 | 72.86 | 73.49 | 47.38 | 23.45 |
| Krum [4] | 94.62 | 83.77 | 81.40 | 78.36 | 49.9 | 38.26 | 87.64 | 86.44 | 83.86 | 81.05 | 36.24 | 21.62 |
| TM [42] | 94.64 | 81.2 | 80.3 | 79.54 | 27.5 | 12.8 | 87.79 | **87.04** | 85.36 | **85.38** | 22.42 | 6.71 |
| Median [42] | 94.68 | 85.86 | 76.64 | 75.42 | 19.86 | 15.42 | 86.92 | 85.34 | 84.32 | 82.36 | 29.83 | 7.92 |
| FLTrust [6] | 12.85 | 9.95 | 9.95 | 9.95 | 9.95 | 9.95 | 10.45 | 5.6 | 4.8 | 4.8 | 4.8 | 4.8 |
| FLAME [32] | 94.58 | 85.26 | 81.2 | 78.24 | 55.46 | 30.6 | 87.39 | 86.31 | 85.71 | 82.38 | 75.17 | 65.63 |
| LoMar [23] | 94.74 | 87.65 | 83.38 | 78.3 | 63.26 | 41.28 | 87.65 | 85.96 | **86.12** | 82.21 | 81.07 | 69.43 |
| FL-Defender [18] | 94.73 | 84.15 | 82.26 | 71.28 | 58.91 | 43.36 | 86.97 | 86.08 | 85.19 | 83.38 | 82.41 | 71.92 |
| n-way | 94.51 | **90.89** | 73.12 | 70.12 | 25.46 | 10.6 | 87.05 | 86.71 | 80.26 | 79.62 | 20.72 | 14.5 |
| FedZZ | 94.76 | 88.95 | **86.24** | **85.23** | **73.56** | **67.43** | 87.84 | 86.57 | 85.55 | 83.61 | **84.17** | **77.74** |

and showed that FedZZ outperforms all the existing techniques across various attack configurations.

> **Answer to RQ2:** *In summary, FedZZ can outperform other existing defense methods under single and multi-client attack settings for both datasets. We observe that FedZZ performs especially well under realistic attack settings ($> 30\%$) in comparison with existing defenses. Overall, FedZZ shows an improvement of $\geq 20\%$, $\geq 10\%$ under 40% attack for CIFAR10 and EMNIST datasets, respectively. Further, FedZZ shows an improvement of $\geq 35\%$, $\geq 11\%$ under 50% attack for CIFAR10 and EMNIST datasets, respectively, demonstrating to be a robust defense.*

## 6.5 Analysis under Varying Degree of non-IID

To assess the effectiveness of our proposed method in handling highly non-IID settings, we conducted experiments on the CIFAR10 dataset with varying degrees of non-IID by using different $\beta$ values (0.1, 0.5, 1, 5, and 10), where lower $\beta$ values resulted in sparsely distributed (unbalanced) data among clients with 0 data in some cases, whereas larger $\beta$ values led to densely distributed (balanced) data with more data per class for each client. To ensure a fair comparison, we evaluated our method against the best performing baselines in each defense category, *i.e.,* Krum, DivFL, LoMar, FL-Defender, and FedAvg. A summary of the results is shown in Figure 6. We present a detailed analysis of corresponding results in Appendix A.1.

> **Answer to RQ3:** *Our results show that FedZZ outperforms existing methods in handling diverse updates, particularly those involving poisonous and highly non-IID updates. Specifically, the accuracy of FedZZ remains less affected by changes in the level of non-IID.*

Under single-client attack conditions, DivFL and Krum perform poorly, as they are designed for good and IID updates, respectively. Under 30% maliciousness, DivFL performs poorly, followed by FedAvg without any defense and Krum. Additionally, as the level of non-IID decreases ($\beta$ increases), all the evaluated methods show a performance improvement.

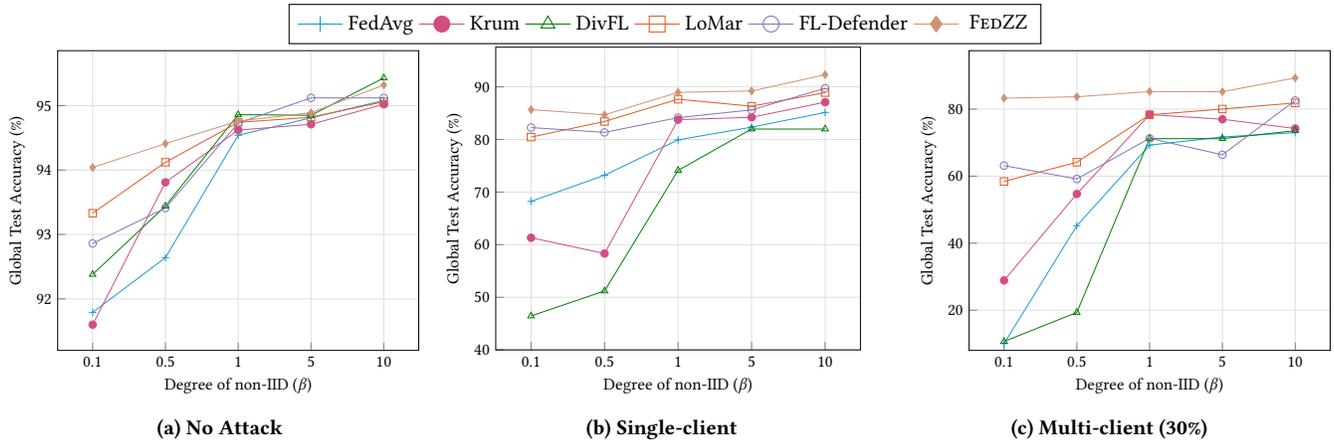

Figure 6: Comparison of GTA of various techniques with different degrees of non-IID using Dirichlet distribution for CIFAR10.

## 6.6 Detection Rate

We evaluate the malicious client detection rate of our approach (FedZZ10, FedZZ5, FedZZ2) against RS, n-way. We calculate the detection rate metric as

$$DR = \frac{|Total\ Dropped\ Malicious\ Updates|}{|Original\ Malicious\ Updates| \times T} \times 100,$$

where $T$ is the number of global epochs. The $DR$ measures the percentage of detections out of all malicious updates over the entire training period of $T$ epochs.

*Single-client Attack:* Figure 9 in Appendix present the results of $DR$ across various configurations of FedZZ under single client attack. As expected, the detection rate of various FedZZ configurations is better than random sampling, demonstrating the effectiveness of the ZBDU technique. The detection rate decreases across all of FedZZ configuration as we increase the calibration interval (Figure 9a v/s 9b and Figure 9c v/s 9d in Appendix). This is expected because of the fewer possible calibrations and, consequently, few chances to determine the appropriate zone map. The trend is the same across both CIFAR (Figure 9a, 9b) and EMNIST (Figure 9c, 9d) datasets. As explained in Section 6.3, decreasing the number of zones tend to increase true positives. We can see this from the higher detection rate of FedZZ5 compared to FedZZ10 as shown in Figure 11 (in Appendix). However, excessively decreasing the number of zones dilutes the effect of the malicious update and consequently results in a lower detection rate, as shown by the lower percentages for FedZZ2 compared to FedZZ5. We observe that as the training progresses, our technique ZBDU learns to detect data poisoning clients, and consequently, detections increase across all configurations.

*Multi-client Attacks:* The results in Figure 7 illustrate the performance of different FedZZ configurations in a multi-client attack scenario, specifically in terms of detection rate ($DR$). These configurations consistently outperform n-way and random sampling methods. The improvement in detection rate is evident as the calibration interval ($\xi$) of FedZZ5 increases for 40% and 50% attack scenarios, as FedZZ5 benefits from a longer period to comprehend the behavior of malicious client updates, particularly in high attack scenarios. However, FedZZ10 and FedZZ2 exhibit varying detection behavior as the calibration interval ($\xi$) is extended due to discrepancies in the number of clients per zone. In this context, n-way aggregation of model updates generates $n$ aggregated models, each containing multiple malicious updates, making it substantially intricate to identify multiple malicious client updates within this framework. As anticipated, FedZZ5 consistently showcases the highest detection rate across numerous configurations, surpassing n-way and random sampling methodologies in multi-client attack scenarios, as evident in both Figure 7a and Figure 7b. The challenge of detecting multiple malicious client updates under the n-way aggregation method is acknowledged, given multiple malicious updates in all aggregated models. Similarly, the challenge of detecting malicious client updates in FedZZ2, which includes two zones with varying proportions of malicious clients, is noted. It's important to note that we avoid comparing our method's detection rate with baselines like DivFL [3], LoMar [23], FL-Defender [18], FLTrust [6], FLAME [32], and Byzantine-robust aggregation techniques, as these approaches involve mechanisms to suppress malicious updates before aggregation, often through scoring, clustering, or weighting model updates. We focus solely on comparing the detection rate of our approach against relevant baseline methods.

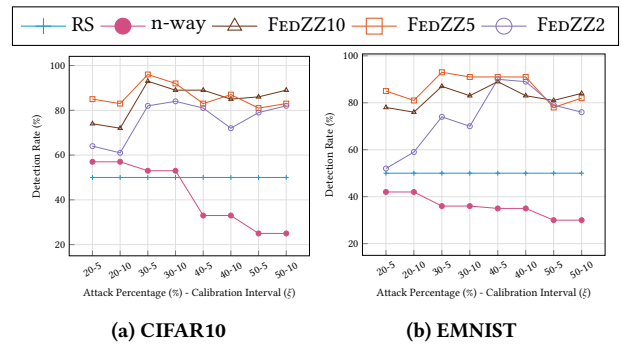

Figure 7: Multi-client attack detection.

*Detection Rate over Training Period.* We present the detection across all epochs for various configurations in Figure 11. Initially, all waveforms have a sharp zigzag shape indicating fewer detections.

> **Answer to RQ4:** *As the training progresses, we can see that detections increase across all configurations indicated by the upper square waveform towards the right. This shows that our ZBDU technique using zone level aggregation can learn to detect data poisoning attacks. The* FedZZ5 *has bigger square regions towards the right, indicating better detection capability compared to other configurations, confirming our intuition.*

### 6.7 False Positive Rate of FedZZ

We evaluate the average false positive rate of our approach (FedZZ10, FedZZ5, FedZZ2) against RS, n-way for $\xi = 5, 10$. Specifically, the percentage of benign updates wrongly discarded as malicious updates. We calculate *AFPR* as

$$AFPR = \frac{|Total\ Dropped\ Benign\ Updates|}{|Total\ Dropped\ Updates| \times T \times 2} \times 100,$$

where $T$ is the number of global epochs. Figure 8 presents *AFPR* results for various FedZZ configurations under single and multi-client attacks for both datasets. *Under single-client attack,* FedZZ2 exhibits lower *AFPR* as the malicious client can be in either of the two zones with a probability of 5% (1/20). Consequently, fewer benign clients are dropped in FedZZ2 compared to FedZZ5 and FedZZ10. The higher *AFPR* of FedZZ5 may be attributed to the trade-off in setting the *cosim* hyperparameter $\alpha = 0.97$, leading to the discard flag $DF = 1$ for some benign updates. Conversely, n-way has malicious updates in all zones except one, resulting in a higher drop of benign updates. For *multi-client attack settings*, we observe a diverse trend for each FedZZ configuration. As expected, n-way with multiple malicious clients in each zone shows higher *AFPR*, followed by FedZZ2 with many poisonous updates distributed in two zones. Furthermore, we observe a common drop in *AFPR* for the 40% attack setting for both datasets. This could be due to the effect of the $\alpha$ hyperparameter for those specific configurations, leading to less *AFPR* than others. However, the same $\alpha = 0.97$ performed differently for other FedZZ configurations. Hence, it is better to consider a *dynamic $\alpha$ setting* or make it a learnable parameter, which we plan to explore in our future work.

### 6.8 Overhead

> **Answer to RQ5:** We found *no observable run-time overhead in using* FedZZ *with the FL system.*

Every calibration round finished within a few seconds as FedZZ uses updates from each of the $\xi$ epochs at the server instead of communicating with the clients. Further, there is no additional communication overhead, as FedZZ follows regular FL protocol by sharing one aggregated model update with each client and receiving one update from each client along with a binary discard flag at the server. On the other hand, FLAME [32] is reported to have an average overhead of 1.67 additional rounds for convergence. FL-Defender [18] commented on their run time overhead and minimal computational overhead at the server. Whereas DivFL [3] works without any overhead. Similarly, other methods also work without the overhead.

## 7 CONCLUSION AND FUTURE WORK

We developed FedZZ, a Precision Guided Approach to identify appropriate client zones effectively. Our evaluation shows that FedZZ effectively maintains good test accuracy close to the maximum possible accuracy under single and multi-client attack settings. Also, it can outperform the existing defense methods under multi-attack settings for the CIFAR10 and EMNIST datasets. In the future, we want to explore the dynamic FedZZ in terms of automatically adapting to a number of zones and stabilization rounds instead of static zones throughout the FL process. Also, we want to apply our FedZZ to other common FL computer vision and NLP tasks.

## ACKNOWLEDGEMENT

This research was supported by in part by the National Science Foundation (NSF) under Grants CNS-2247686, Amazon Research Award (ARA) on "Security Verification and Hardening of CI Workflows" and by Defense Advanced Research Projects Agency (DARPA) under contract number N6600120C4031 and N660012224037. The U.S. Government is authorized to reproduce and distribute reprints for Governmental purposes notwithstanding any copyright notation thereon. Any opinions, findings, conclusions, or recommendations expressed in this material are those of the author(s) and do not necessarily reflect the views of the NSF, Amazon or the United States Government.

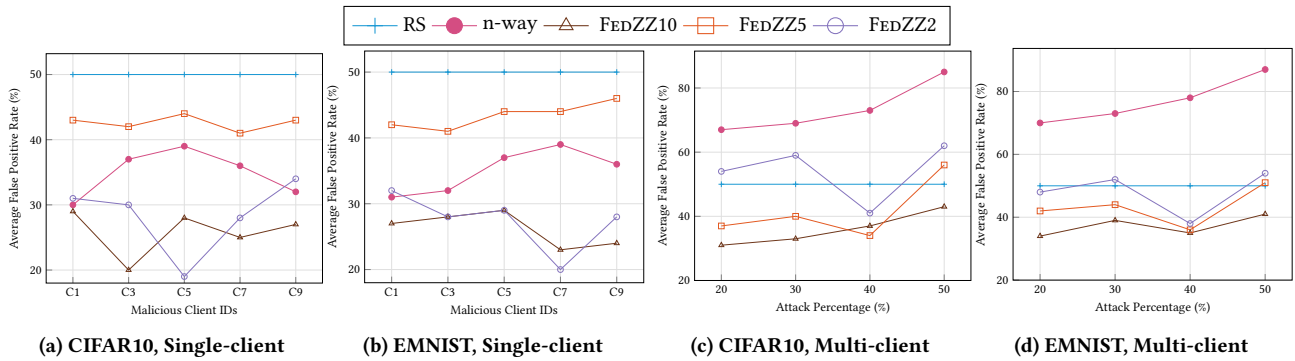

Figure 8: Comparison of average false positive rate of various techniques and configurations on CIFAR10 and EMNIST datasets.

# A APPENDIX

**Table 3: Summary of adopted notations**

| Notation | Definition |
|---|---|
| $C_{dp}$ | Client data poisoning |
| PDAG | Poisoning detection aggregation group |
| ZBDU | Zone based deviating update |
| $n$ | Number of clients |
| $m$ | Number of zones |
| $Z_i$ | $i^{th}$ Zone with $\frac{n}{m}$ clients |
| AZ | Zone level aggregation |
| cosim | Cosine similarity function |
| $D_k$ | $k^{th}$ client local data |
| $w_t^G$ | Global model at epoch $t$ |
| LM | Local model |
| TEE | Trusted execution environment |
| $\alpha$ | Threshold to identify malicious update |
| DF | Discard flag |
| $Z$ | Zones set |
| $\Psi$ | Discriminator map |
| $\xi$ | Calibration interval |
| $Z_{mut}$ | Mutated zone configuration |
| $Z_{new}$ | New zones map |
| IInputs | Interesting inputs |
| $\eta$ | Learning rate |
| $l$ | Loss function |
| $\mathbb{D}_{train}$ | Training data |
| $\mathbb{D}_{test}$ | Test data at the server |
| GTA | Global test accuracy |
| $GTA_{mut}$ | Variable to GTA feedback from server |
| $GTA_{max}$ | Variable to store best maximum GTA |
| DR | Detection rate |
| AFPR | Average false positive rate |
| $\lambda$ | Coefficient in FedAvg function |
| $\beta$ | non-IID Dirichlet data distribution parameter |

## A.1 Effectiveness under Varying Degree of non-IID

Examining diverse non-IID data distributions is crucial for evaluating a defense strategy, providing insights into the influence of the Dirichlet distribution parameter $\beta$ on resulting data characteristics. Analyzing non-IID data distributions is essential for assessing defense effectiveness in varying scenarios, especially as real-world datasets often exhibit heterogeneous distributions. We performed experiments with varied $\beta$ values (0.1, 0.5, 1, 5, and 10) to evaluate our method in highly non-IID settings. Lower $\beta$ values resulted in sparse data distribution among clients, sometimes with zero data, while larger $\beta$ values led to denser distribution. We chose $\beta$=1 as the default value for subsequent experiments. The evaluation included no attack, single-client attack, and multi-client attack scenarios (30% maliciousness) on the CIFAR10 dataset (for brevity), comparing our method to best-performing baselines in each defense category: Krum, DivFL, LoMar, FL-Defender, and FedAvg.

Figure 6 depicts results under no attack, single-client, and multi-client attacks with varying non-IID settings on the CIFAR10 dataset. Our findings showcase FedZZ as a proficient client selection technique under no attack conditions, strategically choosing clients to enhance global test accuracy (Figure 6a). Moreover, our method surpasses others in single-client attack scenarios, even when confronted with a sole poisonous update amid 39 non-IID updates (Figure 6b). By discarding malicious updates and distinguishing between poisonous and highly non-IID updates, our approach improves global model performance. In cases of 30% maliciousness (Figure 6c), where a mix of poisonous and non-IID benign updates exists, our method excels, effectively differentiating between harmful and non-harmful updates, contributing to improved global model performance. *In summary, our proposed method outperforms other methods under diverse updates with respect to poisonous versus non-IIDness and under varying degrees of non-IID and maliciousness.*

## A.2 Effectiveness of Data Poisoning Attacks

**Table 4: Comparison of global test accuracy (%) with existing defenses from different categories under DPA-SLF [37] attack. Bold result indicates the best result for setting.**

| Defense Category | Defense | CIFAR10 | | | EMNIST | | |
|---|---|---|---|---|---|---|---|
| | | No Attack | 1A | 30% | No Attack | 1A | 30% |
| 100% Detection | FL100 | 94.54 | 93.44 | 93.18 | 87.96 | 87.07 | 87.2 |
| Client Selection Techniques | RS | 94.57 | 88.18 | 79.16 | 87.41 | 83.11 | 79.38 |
| | DivFL [3] | **94.86** | 88.12 | 81.19 | 87.26 | 84.05 | 81.64 |
| Aggregation Techniques including Byzantine Robust | FedAvg [29] | 94.54 | 82.03 | 76.27 | **87.96** | 84.76 | 79.68 |
| | Krum [4] | 94.62 | 88.62 | 83.16 | 87.64 | 83.14 | 82.05 |
| | TM [42] | 94.64 | 87.29 | 82.94 | 87.79 | 82.68 | 81.48 |
| | Median [42] | 94.68 | 88.91 | 84.16 | 86.92 | 83.92 | 81.76 |
| | FLTrust [6] | 12.85 | 9.95 | 9.95 | 10.45 | 5.6 | 5.6 |
| Recent Defense Methods | FLAME [32] | 94.58 | 85.26 | 78.24 | 87.39 | 86.31 | 82.73 |
| | LoMar [23] | 94.74 | 87.90 | 84.29 | 87.65 | 83.12 | 82.54 |
| | FL-Defender [18] | 94.73 | 88.83 | 85.36 | 86.97 | 84.29 | 81.64 |
| Ours | n-way | 94.51 | 89.18 | 76.22 | 87.05 | 84.71 | 78.33 |
| | FedZZ | 94.76 | **90.81** | **89.19** | 87.84 | **86.81** | **83.18** |

**MSimBA attack [22]** minimizes the loss of the *most confused class*, resulting in early convergence compared to earlier works [13]. The most confused class is the incorrect class, where the model misclassifies with the *highest* probability, making it an effective attack. **For DPA-SLF [37],** each compromised client flips the labels of their data from true label $y \in [0, C-1]$ to false label $(C-1-y)$ if $C$ is even and to false label $(C-y)$ if $C$ is odd, where $C$ is the number of classes [37]. **For the DPA-DLF attack [37],** we follow [37] by using a surrogate ResNet18 architecture trained on benign data (normal FL model) and flip label $y$ to the least probable label it generates for a given sample. Figures 10a and 10b illustrate the attack success rates of three distinct attacks under our FL setup with no defense and the aforementioned maliciousness levels, namely the single-client attack (1A), as well as the multi-client attack with 20%, 30%, 40%, and 50% maliciousness. The analysis (Figure 10) reveals that the black-box gradient noise data poisoning attack MSimBA outperforms the dynamic and static label flip attacks in the FL setup in terms of attack success rate under no defense. This analysis aids in assessing the effectiveness of proposed FedZZ in reducing the success rate of attacks and improving overall test accuracy.

*In our study, we fixed the malicious client attack ratio as 1.5% (one sample per data batch) for the MSimBA attack [22]. This choice allows us to demonstrate the effectiveness of our approach in detecting low-level attacks with high impact, closely resembling benign updates—a realistic scenario in many applications. Increasing the attack ratio would make the poisoned model significantly differ from benign updates, causing FedZZ to converge faster. We leave this for future work to analyze the impact of varying attack ratios. Additionally, for DPA-SLF [37] and DPA-DLF [37], we flip the labels of all samples in the training dataset due to their low attack success rate.* Furthermore, Table 4 shows the effectiveness of FedZZ against DPA-SLF attack in comparison with other defenses. As expected FedZZ maintains high accuracy in the presence of different poisoning levels. The trend is similar for the DPA-DLF attack.

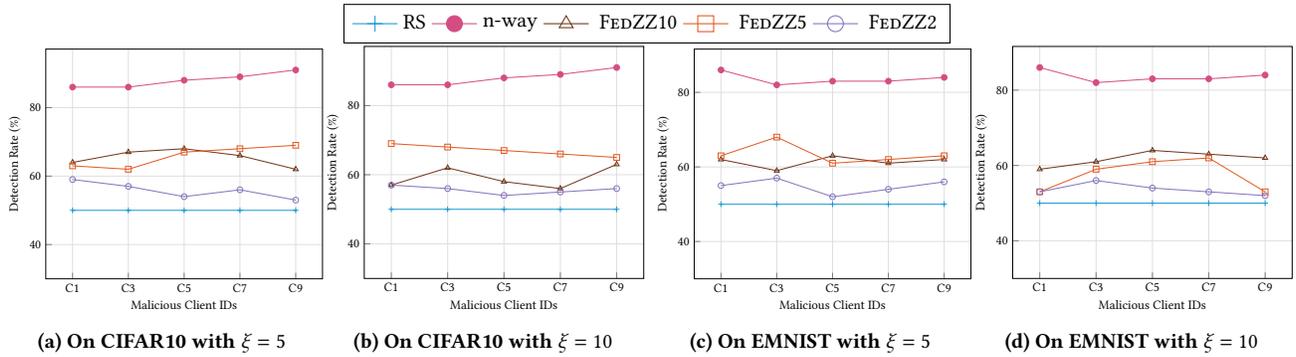

(a) On CIFAR10 with $\xi = 5$　　(b) On CIFAR10 with $\xi = 10$　　(c) On EMNIST with $\xi = 5$　　(d) On EMNIST with $\xi = 10$

**Figure 9: Single client attack detection: Comparison of detection rate of various techniques and configurations on CIFAR10 and EMNIST datasets.**

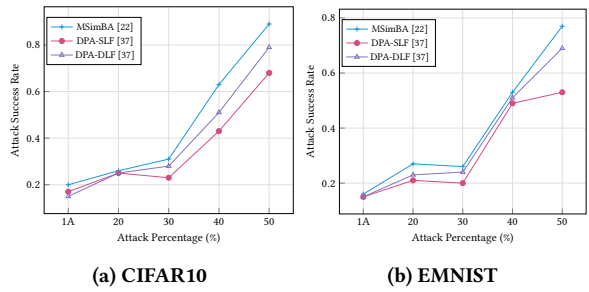

(a) CIFAR10　　(b) EMNIST

**Figure 10: Attack success rate. 1A denotes single-client attack.**

**Table 5: Comparison of global test accuracy (%) of FedZZ with different cosine similarity $\alpha$ values.**

|  | | CIFAR10 | | | | | EMNIST | | | | |
|---|---|---|---|---|---|---|---|---|---|---|---|
|  | Attack | $\alpha = 0.80$ | $\alpha = 0.90$ | $\alpha = 0.95$ | $\alpha = 0.97$ | $\alpha = 1$ | $\alpha = 0.80$ | $\alpha = 0.90$ | $\alpha = 0.95$ | $\alpha = 0.97$ | $\alpha = 1$ |
| No attack | - | 95.05 | 95.33 | 94.98 | **94.76** | 94.31 | 87.24 | 86.32 | 87.25 | **87.84** | 87.71 |
| 1A | MSimBA [22] | 73.39 | 80.32 | 83.47 | **88.75** | 83.21 | 71.63 | 74.26 | 82.33 | **86.57** | 83.11 |
|  | DPA-SLF [37] | 81.23 | 85.09 | 86.01 | **90.81** | 86.54 | 74.80 | 75.40 | 85.37 | **86.81** | 86.28 |
|  | DPA-DLF [37] | 81.49 | 84.34 | 86.25 | **89.12** | 86.75 | 70.06 | 73.83 | 85.88 | **86.38** | 85.67 |
| 20% | MSimBA [22] | 43.62 | 78.69 | 81.98 | **86.24** | 85.26 | 39.68 | 64.43 | 80.96 | **85.55** | 81.28 |
|  | DPA-SLF [37] | 58.37 | 81.23 | 88.70 | **89.94** | 89.59 | 43.09 | 67.73 | 85.65 | **86.14** | 86.35 |
|  | DPA-DLF [37] | 54.36 | 76.17 | 85.42 | **87.43** | 85.80 | 41.65 | 63.92 | **85.98** | 85.55 | 84.31 |
| 30% | MSimBA [22] | 21.03 | 69.45 | 74.63 | 83.96 | **85.23** | 23.43 | 56.98 | 78.42 | **83.61** | 80.64 |
|  | DPA-SLF [37] | 56.75 | 75.91 | 85.47 | **89.19** | 84.54 | 33.44 | 64.59 | 75.96 | **83.18** | 82.13 |
|  | DPA-DLF [37] | 49.12 | 72.44 | 81.29 | **86.15** | 83.88 | 26.82 | 61.28 | 74.18 | 82.74 | **84.78** |
| 40% | MSimBA [22] | 18.96 | 56.43 | 63.24 | **73.56** | 66.12 | 19.23 | 43.32 | 73.29 | **84.17** | 81.23 |
|  | DPA-SLF [37] | 32.38 | 68.29 | **78.63** | 78.63 | 73.24 | 21.71 | 50.56 | 79.68 | **82.31** | 80.11 |
|  | DPA-DLF [37] | 39.35 | 68.88 | 72.47 | **77.21** | 76.34 | 20.26 | 49.67 | **83.33** | 82.62 | 81.85 |
| 50% | MSimBA [22] | 10.24 | 42.61 | 59.43 | **67.43** | 41.36 | 15.26 | 39.91 | 66.28 | **77.74** | 61.13 |
|  | DPA-SLF [37] | 28.44 | 52.59 | 68.96 | **71.43** | 69.91 | 22.99 | 42.35 | 76.27 | **79.62** | 78.51 |
|  | DPA-DLF [37] | 25.49 | 49.31 | 64.23 | **69.42** | 68.29 | 18.10 | 38.98 | 75.11 | **78.45** | 76.67 |

## A.3 Impact of Cosine Similarity

In our evaluation of FedZZ, we varied the cosine similarity threshold parameter ($\alpha$) using five different values (0.80, 0.90, 0.95, 0.97, and 1). Table 5 shows the results obtained for different attack scenarios and settings. We observed that lower $\alpha$ values resulted in true positives being missed, as the similarity threshold was not stringent enough to filter out malicious updates. This allowed even slightly similar malicious updates to pass undetected and be aggregated with the global model. Conversely, a $\alpha$ value of 1 often led to false positives, as even benign updates with slight differences were discarded. This resulted in no updates being considered for aggregation. We found that setting $\alpha$ to 0.97 achieved the best results in most cases. This allowed for the effective filtering of malicious updates while also ensuring that benign updates were not erroneously discarded. Tighter similarity thresholds led to increased accuracy in global testing, as more filtering was possible. Overall, our results demonstrate the importance of selecting an appropriate $\alpha$ value in order to effectively balance the tradeoff between false positives and true positives.

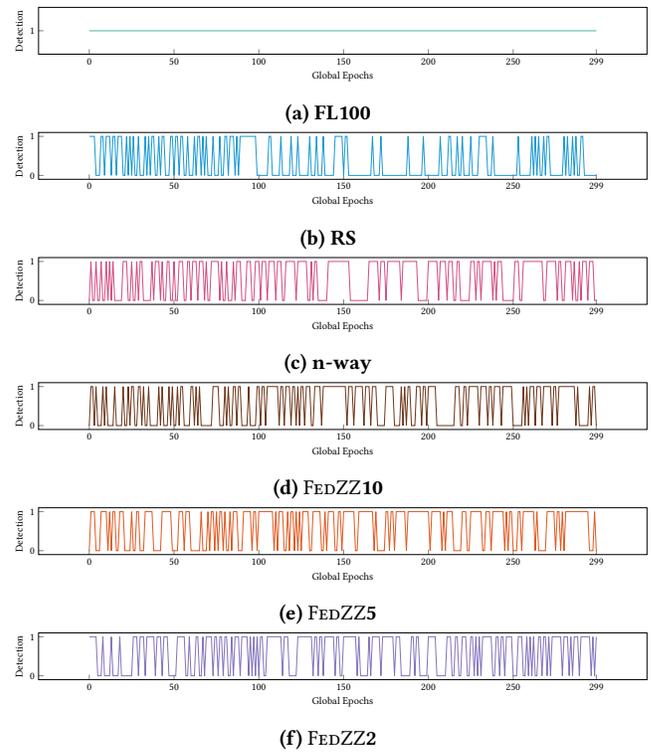

(a) FL100

(b) RS

(c) n-way

(d) FedZZ10

(e) FedZZ5

(f) FedZZ2

**Figure 11: Comparison of detection happening (0/1) of various baselines under single-client attack for 300 global epochs for CIFAR10 dataset.**